\documentclass[floatfix,aps,pre,showpacs]{revtex4}
\pdfoutput=1
\usepackage{graphicx}
\usepackage{amsmath}
\usepackage{amsfonts}
\usepackage{amssymb}
\begin{document}

\title{Three lemmas on the dynamic cavity method}

\author{Erik Aurell}
\affiliation{Department of Computational Biology, AlbaNova University Centre, 106 91 Stockholm, Sweden}
\altaffiliation{Department of Information and Computer Science, Aalto University, Finland}
\altaffiliation{ACCESS Linnaeus Centre, KTH - Royal Institute of Technology, Stockholm, Sweden}
\author{Hamed Mahmoudi}
\affiliation{Department of Information and Computer Science, Aalto University, Finland}

\date{\today}

\begin{abstract}
We study the dynamic cavity method for dilute kinetic Ising models
with synchronous update rules. For the parallel update rule
we find for fully asymmetric models that the dynamic cavity
equations reduce to a Markovian dynamics of the (time-dependent)
marginal probabilities. For the random sequential update rule,
also an instantiation of a synchronous update rule, we find on the
other hand that the dynamic cavity equations do not reduce to
a Markovian dynamics, unless an additional assumption of
time factorization is introduced. For symmetric models we
show that a fixed point of ordinary Belief propagation is
also a fixed point of the dynamic cavity equations in the
time factorized approximation. For clarity, the conclusions
of the paper are formulated as three lemmas.  
\end{abstract}


\maketitle 

\section{Introduction}
For diverse applications to information theory, artificial intelligence
and other fields, as well as in the physics of dilute spin glasses, 
much attention has been given over the last decade to a class
of distributed computational schemes known as
iterative decoding, Belief Propagation (BP) or the 
cavity method~\cite{mezard2,yedidia}. These methods
determine the marginals of Markovian random fields,
where the dependency structure is encoded by a factor graph.
The method is exact if the factor graph is a tree, and 
often very accurate if the factor graph is locally tree-like.
The prime examples of such locally tree-like graphs graphs are
on random graphs or random hyper-graphs, which underlie, for instance,
LPDC codes, random graph coloring and
and random $k$-satisfiability, and the
dilute Sherrington-Kirkpatrick spin glass.

A general feature of most applications to date of these schemes
is that they target marginals of Boltzmann-Gibbs measures,
which describe physical systems in equilibrium. Such measures are 
also the stationary state of (families of) 
Monte Carlo (MC) schemes, 
where the update rules obey detailed balance. The main 
advantage of BP is then that it is (typically) many times faster than 
MC, and therefore the preferred choice when marginals of 
Boltzmann-Gibbs measures have to be computed 
both accurately and efficiently.   

A synchronous update Monte-Carlo scheme to simulate \textit{e.g.} 
an Ising spin glass can be visualized as a tower of 
variable sets, where 
each horizontal layer represent the spins at some time $t$, and where
the links between the layers encode the dependences in the update 
rules. Such a description is not limited to equilibrium physics,
but extends to update rules which do not obey detailed balance.

The question then naturally poses itself whether a distributed
computational scheme can be found which computes the marginal
distributions of such factor graphs. Kanoria and
Montanari in~\cite{montanari} showed that this is the case
for \textit{majority dynamics} on trees, while Neri and Boll\'e
in~\cite{bolle}
showed that given an assumption which we will call 
\textit{time factorization} the (asymmetric, non-equilibrium) 
Ising spin glass with parallel update rule also leads to a BP-like
scheme. In~\cite{us} we 
extended the Neri-Boll\'e approach to a sequential update rule, and 
showed that it gives indeed in many cases very accurate predictions of
the marginals of stationary non-equilibrium states.

These results on \textit{dynamic cavity method} are, we believe, 
quite important, as potentially pointing to a new class of general, 
accurate and efficient approximation schemes in non-equilibrium systems.
They therefore deserve further study, outlining when and how
they work, and when they don't. In this contribution we will 
address the following aspects of the systems studied 
in~\cite{bolle} and~\cite{us}: 
\textit{(i)} does the dynamic cavity
reduce to a Markovian dynamics
if the underlying graph is fully asymmetric?
\textit{(ii)} is there a difference depending on which
update rule (parallel or sequential) is used? 
\textit{(iii)} what is the relation between the dynamic
cavity method and ordinary BP if the underlying graph is 
symmetric, and hence describes an equilibrium system?

The answers to the first two questions, which we formulate as 
Lemma~1 and Lemma~2 below, are that for fully asymmetric graphs and 
the \textit{parallel} update rule, the dynamic cavity equations
do reduce to a Markovian dynamics, without additional assumptions, 
but for the \textit{sequential} update rule this is not so.
We remark that for tractability the dynamic cavity equations
must be reduced to a Markovian dynamics, as otherwise one would
have to keep track of the whole history in a simulation.
Therefore, Lemma~1 and Lemma~2 also mean that for parallel
updates on a fully asymmetric graph, the reduced cavity equations
are in a certain sense exact, and hold both for transients and
stationary states, while for sequential updates this is not so.
Indeed, in~\cite{us} we only found good agreement between
the reduced cavity equations and MC under sequential updates
for the stationary states, but not for transients. 
The answer to the third question, which was already stated in~\cite{bolle},
is that a fixed point of BP on an equilibrium Ising model can 
be extended to a fixed point of dynamic cavity method for the same model. 
As no proof of this result has appeared in the literature (to the best of
our knowledge), we include it here as Lemma~3. 

The paper is organized as follows:
in section~\ref{sec:recap} we recall the salient features
of dynamic cavity method applied to dilute Ising spin systems;
in section~\ref{sec:asymmetric} we consider fully asymmetric
systems and state and prove Lemma~1 and Lemma~2; and in
section~\ref{sec:symmetric} we consider symmetric 
systems and state and prove Lemma~3. In section~\ref{sec:conclusion}
we briefly summarize our results. References to the 
earlier literature, especially as pertaining to other
methods to analyze the systems under consideration, are
given where appropriate throughout the paper.

\section{Microscopic dynamics for asymmetric dilute Ising models}
\label{sec:recap}
The asymmetric dilute Ising model is defined over 
a set of $N$ binary variables $\vec{\sigma}=\{\sigma_1,\ldots,\sigma_N\}$,
and an asymmetric graph $G=(V,E)$ where
$V$ is a set of $N$ vertices, and $E$ is a set of directed edges.
We use the notation \textit{fully asymmetric} when 
if there is an edge $(v_i,v_j)$ there is no edge $(v_j,v_i)$, and
\textit{symmetric} when if there is an edge $(v_i,v_j)$ there
is also an edge $(v_j,v_i)$. A symmetric diluted Ising model
is hence here a special case of an asymmetric dilute Ising model.
To each vertex $v_i$ is associated a binary variable $\sigma_i$.
The graphs $G$ are taken from random graph ensembles with
bounded average connectivity $c$. 

The microscopic description of the dynamics of such system 
with an synchronous update rule is a Markovian dynamics for the evolution of the joint probability distribution
\begin{equation}
p(\vec{\sigma}(0),...,\vec{\sigma}(t)) = \prod_{s=1}^t\, W(\vec{\sigma}(s)\, |\, \vec{h}(s) ) p(\vec{\sigma}(0)) 
\label{eq:dynamics}
\end{equation}
where the transition matrix $W$ depends on local fields associated to spins denoted by $\vec{h}$
\begin{equation}
h_i(s) = \sum_{j\in \partial i} J_{ji} \sigma_j(s-1) + \theta_i(t).
\end{equation}
and the local fields determine jump rates
\begin{equation}
w_i(\sigma_i(t)|h_i(t)) = \frac{1}{1+\exp(2\beta\sigma_i(t)h_i(t))}
\end{equation}
In a synchronous update rule, one, some or all the spins are updated in each time step. We will here consider the two extreme cases, where either all spins are updated (\textit{parallel update rule}), or where just one randomly chosen spin is updated (\textit{sequential update rule}):
\begin{eqnarray}
\label{eq:update}
W(\vec{\sigma}(t)\,|\, \vec{h}(s)) =
\left\{
\begin{array}{c cl}
  \prod_{i=1}^N\, w_i(\sigma_i(t)\,|\,h_i(t)) && {\rm parallel \,\,\,\, update} \\
 1/N\sum_{i}\, \prod_{j\neq i}\delta_{\sigma_j(t),\sigma_j(t-1)}\, w_i(\sigma_i(t)\,|\,h_i(t)) &&{\rm sequential \,\,\,\, update}
\end{array}
\right.
\end{eqnarray}
We note that the sequential update rule is not the same as
asynchronous updates (Glauber dynamics), because the decisions of
which spin is chosen and whether/whether not/ to flip that spin are
here taken in the opposite order.

Equation~(\ref{eq:dynamics}) can be marginalized over one spin $i$,
or over the neighborhood of that spin $\partial i$.
The probability of observing a history of a single spin 
then follows a self-consistency equation, which for the parallel 
update reads
\begin{eqnarray}
p_i(\sigma_i(0),...,\sigma_i(t) \,|\, \theta_i(0),...,\theta_i(t)) = p_i(\sigma_i(0))\sum_{\vec{\sigma}_{j\in \partial i}(0),\ldots,\vec{\sigma}_{j\in \partial i}(0)}&& p_{\partial i}(\sigma_{j\in \partial i}(0),\ldots,\sigma_{j\in \partial i}(t)\,|\,\sigma_i(0),...,\sigma_i(t)) \nonumber\\
&&\prod_{s=1}^t w_i(\sigma_i(s)|h_i(s))
\label{eq:par_dynamic}
\end{eqnarray}
and for the sequential updates
\begin{eqnarray}
p_i(\sigma_i(0),...,\sigma_i(t) \,|\, \theta_i(0),...,\theta_i(t)) = p_i(\sigma_i(0))&&\sum_{\vec{\sigma}_{j\in \partial i}(0),\ldots,\vec{\sigma}_{j\in \partial i}(0)}  p_{\partial i}(\sigma_{j\in \partial i}(0),\ldots,\sigma_{j\in \partial i}(t)\,|\,\sigma_i(0),...,\sigma_i(t)) \nonumber\\
&&\prod_{s=1}^t \left(\frac{1}{N} w_i(\sigma_i(s)|h_i(s)) + (1-\frac{1}{N}) \delta_{\sigma_i(s),\sigma_i(s-1)}\right)\,\,\, .
\label{eq:seq_dynamic}
\end{eqnarray}
The variables appearing in the conditional probability
$p_{\partial i}$ above are the histories of the cavity spins, 
and the two equations can be considered output equations for the dynamic
cavity method. In the Belief propagation approximation the 
histories of the cavity
spins are independent conditional on the history of spin $i$, which means
\begin{equation}
p_{\partial i}(\sigma_{j\in \partial i}(0),\ldots,\sigma_{j\in \partial i}(t)\,|\,\sigma_
i(0),...,\sigma_i(t)) = \prod_{j\in \partial i} \mu_{j\to i}(\sigma_j(0),\ldots,\sigma_j(t)\,|\, \sigma_i(0)\ldots\sigma_i(t))
\label{eq:bp}
\end{equation}
where $\mu$ denotes a marginal probability of the history of cavity spin $j$, conditioned on the history of spin $i$ (dynamic BP messages).
Note that in Eq.~\ref{eq:seq_dynamic} and Eq.~\ref{eq:par_dynamic} the set of spins contributing in the trajectory of spin $i$
are those with a directed incoming link to spin $i$. 

The evolution of marginal probability at time $t$ can then be obtained by summation over the past history, {\it i.e}, $p_i(\sigma_i(t)) = \sum_{\sigma_i(0),\ldots,\sigma_i(t)}\, p_i(\sigma_i(0),\ldots,\sigma_i(t))$.
It is straightforward to verify that in general the evolution of $p_i(\sigma_i(t))$ requires information from the whole past history and therefore 
is a non-Markovian process. 
A main result of~\cite{bolle} and~\cite{us} was that by a 
a further assumption of time factorization of the dynamic BP messages,
the evolution of $p_i(\sigma_i(t))$ is a Markov chain of order~2
(\textit{i.e.} the evolution
requires information on one and two time steps earlier).
Intuitively, it may be argued that for fully asymmetric models,
where if dynamic BP messages go out they don't come back unless
going around a long loop in the graph, the $p_i(\sigma_i(t))$
should obey Markovian dynamics, without the assumption of 
time factorization. In the following section we will show that
this indeed is the case for parallel updates, where time
factorization always holds -- but it is not true for 
sequential updates.

We end this section by a remark on random graph ensembles.
In~\cite{us} we followed the parameterization of~\cite{coolen}
using a connectivity matrix $c_{ij}$, where $c_{ij}=1$ if there is
a link from vertex $i$ to vertex $j$, $c_{ij}=0$ otherwise, and matrix elements
$c_{ij}$ and $c_{kl}$ are independent unless $\{kl\}=\{ji\}$.
In this parameterization the random graph is specified by marginal (one-link) distributions
\begin{equation}
p(c_{ij}) = \frac{c}{N}\delta_{1,c_{ij}} + (1 - \frac{c}{N})\, \delta_{0,c_{ij}}\,\,\,.
\label{eq:connec}
 \end{equation}   
and the conditional distributions
\begin{equation}
p(c_{ij}\,|\,c_{ji}) = \epsilon \delta_{c_{ij},c_{ji}} + (1-\epsilon) \, p(c_{ij})\,\,\,.
\label{eq:asym}
\end{equation} 
In this model the average degree distribution is given by $c$,
and the asymmetry is controlled by $\epsilon\in[\,0,1\,]$.
The results given below describe $\epsilon=0$ 
(Lemma~1 and Lemma~2, section~\ref{sec:asymmetric})
and $\epsilon=1$ (Lemma~3, section~\ref{sec:asymmetric}). 
In the first case the analogy is however only exact in the 
limit of large system size.

\section{Fully asymmetric networks}
\label{sec:asymmetric}
In this section we assume
fully asymmetric diluted Ising models such that if spin $i$ is 
connected to spin $j$ then spin $j$ does not connect back to spin $i$.
This property simplifies the evolutionary equations of single site 
probability because influences (through interactions) do not return.

We consider the two update rules separately.
\subsection{Fully asymmetric models -- parallel update}
{\bf Lemma~1} \,\,\,\,\,\,\,\,The following recursive equation holds for the fully asymmetric networks
\begin{equation}
p(\sigma_i(t)) = \sum_{\vec{\sigma}_{j\in \partial i}(t-1)}\, p(\vec{\sigma}_{j\in \partial i}(t-1)) \, \frac{e^{\beta \sigma_i(t)\,h_i(t)}}{2\, \cosh(\beta h_i(t))}
\end{equation}
The lemma hence states that the evolution of single site distribution in parallel update 
follows a Markovian process when the network is fully asymmetric. Therefore 
at each iteration we only need to have information about one iteration step before. \\

{\bf Proof}
The proof is a straight-forward consequence of the definitions. 
For this update rule the marginal probability of spin 
$i$ at time $1$ (after the first
update) is $p_i(\sigma_i(1)) = \sum_{\sigma_{j\in \partial i}} p_{\partial i}(\sigma_{\partial i}(0)) w_i(\sigma_i(1)|h_i(1)$, where the marginal $p_{\partial i}(\sigma_{\partial i}(0))$ is given by the initial conditions. 
For the marginal probability after $t$ steps we have 
\begin{equation}
p(\sigma_i(t)) = \sum_{\sigma_i(0),...,\sigma_i(t-1)}\sum_{\sigma_{\partial i}(0),...,\sigma_{\partial i}(t-1)}\, p(\sigma_{\partial i}(0),...,\sigma_{j\in \partial i}(t-1)) \, \prod_{s=1}^t \frac{e^{\beta \sigma_i(s)\,h_i(s)}}{2\, \cosh(\beta h_i(s))}
\end{equation}
Since the neighborhood $\partial i$ denotes the set of spins 
connected to spin $i$ by a link incoming to $i$, and since in fully
asymmetric models such spins will not be connected to $i$
by a link outgoing from $i$, the probability 
$p_{\partial i}(\sigma_{\partial i}(0),...,\sigma_{j\in \partial i}(t-1))$
is in this case independent of the history of spin $i$.
We can therefore sum over time, and the 
only remaining term is the joint probability distribution 
of the cavity spins one update before $t$. 
\begin{equation}
p(\sigma_i(t)) = \sum_{\vec{\sigma}_{j\in \partial i}(t-1)}\, p(\vec{\sigma}_{j\in \partial i}(t-1)) \, \frac{e^{\beta \sigma_i(t)\,h_i(t)}}{2\, \cosh(\beta h_i(t))}
\end{equation}
{\bf End of proof}
\\\\
It is worth pointing out that the above conclusion is true in general 
and does not rely on the Belief propagation approximation. Indeed, we have not used the BP approximation in the calculations above. The corresponding output equation for dynamic BP reads 
\begin{equation}
p(\sigma_i(t)) = \sum_{\vec{\sigma}_{j\in \partial i}(t-1)}\, \prod_{j\in\partial i}\mu_{j\to i}(\sigma_{j\in \partial i}(t-1)) \, \frac{e^{\beta \sigma_i(t)\,h_i(t)}}{2\, \cosh(\beta h_i(t))}
\label{eq:bp_asym_par}
\end{equation}
The dynamic BP messages themselves obey the following recursion equations 
\begin{equation}
\mu_{i\to j}(\sigma_i(t)) = \sum_{\vec{\sigma}_{k\in \partial i\setminus j}(t-1)}\, \prod_{k\in\partial i\setminus j}\mu_{k\to i}(\sigma_{k}(t-1)) \, \frac{e^{\beta \sigma_i(t)\,h_i^{(j)}(t)}}{2\, \cosh(\beta h_i^{(j)}(t))}
\label{eq:bp_asym_mess_par}
\end{equation}
where $h_i^{(j)}$ is the effective field on spin $i$ in the cavity graph, $h_i^{{j}} = \sum_{k\in\partial i\setminus j} J_{ki} \sigma_k(t-1) + \theta_i(t)$.
\subsection{Fully asymmetric models -- sequential update}
{\bf Lemma~2} \,\,\,\,\,\,\,\,
For sequential updates,
the time evolution of marginal probability distribution 
does not follow a Markovian process, and generally depends on the whole history.\\\\
{\bf Proof}
The proof proceeds by showing that a reduction analogous to
the proof of Lemma~1 above does not take place for sequential updates. 
The marginal probability distribution after one step is given by 
\begin{equation}
p_i(\sigma_i(1)) = \sum_{\sigma_i(0)}\sum_{\sigma_{j\in \partial i}(0)} p(\sigma_{j \in \partial i}(0)) \left(\frac{1}{N}\,w_i(\sigma_i(1)|h_i(1))+(1-\frac{1}{N}) \delta_{\sigma_i(0),\sigma_i(1)}\right) p_i(\sigma_i(0))
\end{equation}
Performing the summation over $\sigma_i(0)=\{-1,1\}$ will split this equation into two parts
\begin{equation}
p_i(\sigma_i(1)) = \frac{1}{N}\sum_{\sigma_{\partial i}(0)}\, p(\sigma_{j \in \partial i}(0))w_i(\sigma_i(1)|h_i(1)) + (1-\frac{1}{N}) p^{(0)}_i(\sigma_i(1))
\end{equation}
The last term is the probability distribution $p^{(0)}_i(\sigma_i(1))$ over spin $i$ at time $0$, but taking as argument the 
value of spin $i$ at time $1$. It is clear that this term is problematic, and we will show that this problem does not go away.
After two iterations we have
\begin{eqnarray}
p_i(\sigma_i(2)) &=& \frac{1}{N^2} \sum_{\sigma_{\partial i}(1)} p(\sigma_{\partial i}(1)) w_i(\sigma_i(2)|h_i(2)) \nonumber\\
&+& \frac{1}{N}(1-\frac{1}{N}) \sum_{\sigma_{\partial i}(1)}  p(\sigma_{\partial i}(1)) w_i(\sigma_i(2)|h_i(2))  \nonumber\\
&+& \frac{1}{N}(1-\frac{1}{N}) \sum_{\vec{\sigma}_{\partial i}(1)} p(\sigma_{\partial i}(1)) w_i(\sigma_i(2)|h_i({\bf 1}))  \nonumber\\
&+&(1-\frac{1}{N})^2 p^{0}(\sigma_i(2))
\end{eqnarray}
The first two terms partially cancel, but 
that is all.
Therefore, the equation for the evolution of marginal probabilities at iteration step $t$ contains all sequences of possible 
update series
\begin{eqnarray}
p_i(\sigma_i(t)) &=& \frac{1}{N^t}   \sum_{\sigma_{\partial i}(t-1)} p(\sigma_{\partial i}(t-1)) w_i(\sigma_i(t)|h_i(t)) \nonumber\\
&+& (1-\frac{1}{N})^t    p^{0}(\sigma_i(t)) \nonumber \\
&+& \sum {\cal F}({\rm 0,...,t-1})
\end{eqnarray}
where the first term corresponds to the case where spin $i$ has been updated all the time, the second term is for the case where 
it has never been updated and the last term stands for all permutation of different update trajectory in which none of the first two cases happen.\\ 
{\bf End of proof}.

\section{Symmetric networks}
\label{sec:symmetric}
We begin by introducing the time factorization ansatz for
models which are not fully asymmetric. In both cases, these
amount to the assumption
\begin{equation}
\mu_{j\to i}(\sigma_j(0),\ldots,\sigma_j(t)\,|\, \sigma_i(0)\ldots\sigma_i(t-1))
= \mu_{j\to i}^{(0)}(\sigma_j(0)) \prod_{s=1}^t \mu_{j\to i}^{(s)}(\sigma_j(s) \,|\, \sigma_i(s-1))
\end{equation}
and lead to
respectively 
\begin{equation}
p^{t}_{i}(\sigma_i(t)) = \sum_{\vec{\sigma}_{\partial i}(t-1)}\,\prod_{k\in\partial i} \mu^{t-1}_{k\to i}(\sigma_k(t-1))\,\,w_i(\sigma_i(s)\,|\,h_i^{(j)}(s)) p^{t-2}_{i}(\sigma_i(t-2))
\label{eq:bp_mess_syn_fully_asym}
\end{equation}
for parallel updates, and 
\begin{eqnarray}
p^t_{i}(\sigma_i(t)) =\frac{1}{N} p^{t-1}_{i}(\sigma_i(t)) &+& (1-\frac{1}{N}) \sum_{\sigma_i(t-2),\vec{\sigma}_{\partial i\setminus j}(t-1)} \prod_{k\in\partial i}\,\mu^{t-1}_{k\to i}(\sigma_k(t-1)\,|\,\theta_i^{(j)}(t-2))\nonumber\\
&&w_i(\sigma_i(t)\,|\,h_i(t))\,p^{t-2}_{i}(\sigma_i(t-2))
\label{eq:bp_mess_asyn_fully_asym}
\end{eqnarray} 
for sequential updates.
The numerical results reported in~\cite{us} are based on these equations. 
For fully asymmetric models, Eq.~\ref{eq:bp_mess_syn_fully_asym} reduces to Eq.~\ref{eq:bp_asym_par}. 

In the following we will discuss the fixed points of 
(\ref{eq:bp_mess_syn_fully_asym}) and 
(\ref{eq:bp_mess_asyn_fully_asym}) -- which are obviously
the same -- for symmetric networks, and 
show that the fixed points of
ordinary Belief propagation also solve these equations.
This property was stated in~\cite{bolle} and, from the viewpoint
of generating functional analysis, already seven years ago in~\cite{coolen}.
A proof has however to our knowledge not appeared based on dynamic cavity formalism..\\
{\bf Lemma~3}
In stationary state, the ordinary BP equations satisfy Eq.~\ref{eq:bp_mess_syn_fully_asym} and Eq.~\ref{eq:bp_mess_asyn_fully_asym}.\\
{\bf Proof:}
Introducing the usual cavity fields for the dynamic messages, $\mu^t_{i\to j}(\sigma_i(t)) = \frac{\beta\,u_{i\to j}(t)\,\sigma_i(t)}{2\cosh(u_{i\to j}(t))}$ we can rewrite Eq.~\ref{eq:bp_mess_syn_fully_asym}
in terms of cavity fields. They fulfill the following equations
\begin{eqnarray}
u_{j\to i}^{t} + \theta_j = \frac{1}{\beta}\sum_{\sigma_j^{t}}\sigma_j^{t} \log\left\{\sum_{\vec{\sigma}_{\partial j\setminus i}^{t-1},\sigma_j^{t-2}}\frac{\exp[\beta \sigma_j^{t}(\sum_{k\in \partial j\setminus i}J_{ki}\sigma_k^{t-1} + \theta_j)]}{2\cosh[\beta (\sum_{k\in \partial j\setminus i}J_{ki}\sigma_k^{t-1} + \theta_j)]}\right.\nonumber\\
\left. \prod_{k\in \partial j \setminus i}\frac{\exp[\beta(	\sigma_k^{t-1} u_{k\to j}^{t-1} + \sigma_{k}^{t-1}\sigma_j^{t-2} J_{kj})]}{2\cosh[\beta(u_{k\to j}^{t-1} + \sigma_j^{t-2} J_{kj})]}\,\frac{\exp[\beta \sigma_j^{t-2}u_{j\to i}^{t-2}]}{2\cosh[\beta \sigma_j^{t-2}u_{j\to i}^{t-2}]}\right\}
\label{eq:cavity}
\end{eqnarray}
where the variables are indexed by spin number and time.
These equations can be simplifies using the following two 
well-known formula 
\begin{equation}
2\cosh[\beta(u + J \sigma)] = c(u,J) \exp(\beta V(u,J) \sigma)
\label{eq:cosh}
\end{equation}
where
\begin{eqnarray}
c(u,J) = 2\frac{\cosh(\beta u)\cosh(\beta J)}{\cosh(\beta V(u,J))} \\
V(u,J) = \frac{1}{\beta} {\rm atanh}[\tanh(\beta J)\tanh(\beta u)]
\end{eqnarray}
The factorized normalization term in the Eq.~(\ref{eq:cavity}) then reduces to 
\begin{equation}
 \prod_{k\in \partial j \setminus i} \frac{1}{2\cosh[\beta(u_{k\to j}^{t-1} + \sigma_j^{t-2} J_{kj})]} = \left(\prod_{k \in \partial j \setminus i} \frac{1}{c(J_{jk},u_{k\to j}^{t-1})}\right ) \exp(\beta \sigma_j^{t-2}\sum_{k\in \partial j \setminus i}V(J_{ jk},u_{k\to j}^{t-1}))
\label{eq:factorized-normalization}
\end{equation}
Equation~(\ref{eq:cavity}) using (\ref{eq:factorized-normalization})
is not ordinary BP equations,
but we can show that it admits fixed points of ordinary BP as a 
fixed point. 
We first assume that Eq~(\ref{eq:cavity}) is at a fixed point,
so that the time indices can be ignored. Then we interpret 
the messages in Eq.~(\ref{eq:cavity}) as ordinary BP messages,
and compare to the BP fixed point equations for the diluted Ising spin glass:
\begin{equation}
\sum_{k\in \partial j \setminus i}V(J_{ jk},u_{k\to j}^{t-1}) =\sum_{k\in \partial j \setminus i} \frac{1}{\beta} {\rm atanh}[\tanh(\beta J_{kj})\tanh(u_{k\to j})] = \theta_j + u_{j\to i}
\label{eq:BP-Ising-fixed-point}
\end{equation}
It is seen that the solutions to Eq~(\ref{eq:BP-Ising-fixed-point})
are then also solutions to Eq~(\ref{eq:cavity}).\\
{\bf End of proof}.

The ordinary BP equations are not necessarily the only solution to the dynamic cavity equations in the time-factorization approximation.
It would be of interest to investigate whether the temperature in which ordinary BP starts to fail coincides with the temperature where dynamic cavity equations do not converge to a fixed point. We plan to return to this point in a future
contribution.

\section{Conclusion}
\label{sec:conclusion}
The dynamic cavity method is a way to compute (approximately) 
marginals of non-equilibrium states. It has recently been shown by us 
and others to be exact in certain cases, and surprisingly accurate
in a larger class of models. Since computing marginals of non-reversible
Markov chains is a rather general problem, it is clearly important
to outline when these methods can be expected to be accurate, and/or
exact. In this paper we have looked at these questions for 
fully asymmetric models, for parallel and for sequential updates,
and for symmetric models. A major open problem at the moment is
if this approach can be extended from synchronous to asynchronous
update rules.

We end by a short discussion where these methods could be useful.
First, non-equilibrium physical systems live in finite-dimensional
space, and have (on the lattice) factor graphs with many short
loops. This is therefore not a setting where the dynamic cavity
method would be expected to be competitive.
Applications should instead be sought in systems 
(social, technological, biological,$\ldots$) which can 
reasonably be modelled 
by sparse random graphs or hyper-graphs. One such 
application could be
describing bargaining processes to reach agreement 
through local interactions, as in the majority game for consensus
investigated in~\cite{montanari}. 
Another could be describing networks of queues, which,
in contrast to standard queueing theory, do not obey a partial
balance condition~\cite{Kellybook}. Models of this kind were
investigated numerically some time ago to determine blocking
probability in certain types of mobile communication 
systems~\cite{vazquez-abad2002}, and dynamic cavity method
could be of relevance to speed up such estimations.
A third could finally be to improve upon 
network inference algorithms of the ``kinetic Ising'' 
type~\cite{hongli1,yasserhertz,yasser_tap,hongli2}
through more accurate estimates of the direct problem.

\section*{Acknowledgement}
We thank Silvio Franz, Izaak Neri and Lenka Zdeborov\'a for useful 
discussions, and the Kavli Institute of Theoretical Physics China
for hospitality. The work was supported by the
Academy of Finland as part of its Finland Distinguished
Professor program, project 129024/Aurell, and
in part by the Project of Knowledge Innovation Program (PKIP) of Chinese Academy of Sciences, Grant No. KJCX2.YW.W10.

\end{document}